# Plane-wave Density Functional Theory Study of the Electronic and Structural Properties of Ionized and Neutral Small Gold Clusters

**Byeong June Min[*], Won Chul Shin[†], and Jae Ik Park**

*Department of Physics, Daegu University, Kyungsan 712-714, Korea*

We studied the structural and the electronic properties of ionized and neutral small gold clusters ($Au_n$, $Au_n^+$, $Au_n^-$, n = 2 ~ 8) via plane wave pseudopotential calculations. All except $Au_7^-$ favors one-dimensional zigzag structure or two-dimensional arrangement of triangles. The HOMO-LUMO (highest occupied molecular orbital – lowest unoccupied molecular orbital) gap, the ionization energy, and the electron affinity exhibit even-odd variation as a function of the cluster size.

[*]Email: bjmin@daegu.ac.kr

Fax: +82-53-850-6439, Tel: +82-53-850-6436

[†]Current address: Department of Physics, Pohang Institute of Science and Technology, Pohang, KOREA.

Small clusters distinguish themselves from their solid state counterpart in that their structural, electronic, and magnetic properties sensitively depend on their size and on their charge states. Such diversity opens up a wide range of possible applications. For example, gold clusters and their ions are noted for their participation in catalytic reactions, although gold metal is known as a noble metal for its very low chemical activity. Gold nanoparticles are now being used as the vehicle for targeted drug delivery system. With the progress of nanotechnology, the need to understand their physical and chemical properties at an atomic level is growing.

Small Au clusters and their ions exhibit interesting characteristics. Their mass spectra obtained from secondary ion mass spectroscopy (SIMS) and liquid metal ion source (LMIS) show a sawtooth-shaped behavior, that is, an even-odd alternation of the stability [1,2]. We investigate the geometries and electronic properties of these small clusters and their ions via plane-wave density functional theory calculations.

The calculations were performed by using the ABINIT package [3]. A norm-conserving scalar-relativistic pseudopotential [4] is used. Exchange correlation energy was described by the Perdew-Burke-Ernzerhof parameterization within the generalized gradient approximation (PBE-GGA) [5]. Plane-wave cutoff energy of 1088 eV is used. A cubic box of 13.25 Å each side is introduced for periodic boundary condition. The Broyden-Fletcher-Goldfarb-Shanno (BFGS) quasi-Newtonian minimization scheme [6] is used in the structure optimization.

The Au dimer and its anion have served as a reference system for the photoelectron spectroscopy [7-9]. Au dimer ion has also drawn attention for its catalytic activity [10]. In our calculation, neutral Au dimer has a bond length of 2.551 Å, vibrational frequency 173.4 cm$^{-1}$, and the dissociation energy 2.218 eV, in good agreement with 2.4715 Å, 194 cm$^{-1}$, and 2.290 eV, respectively, from experiments [8]. The Au$_2$ anion has a bond length of 2.658 Å, vibrational frequency 134.2 cm$^{-1}$, and dissociation energy 1.920 eV (with respect to one Au atom and one Au$^-$ anion). These compare well with 2.587 Å, 153 cm$^{-1}$, and 1.937 eV, respectively, from a recent photoelectron spectroscopy experiments [8]. The vertical ionization energy of the Au dimer is estimated by considering the total-energy of Au$_2^+$ without further geometrical relaxations to simulate the Franck-Condon principle. When the monopole correction term to the total-energy is taken into account, the vertical ionization energy of Au$_2$ is 9.520 eV compared to 9.50 eV from experiment [9]. In comparison, the adiabatic ionization energy is 9.487 eV. These results are very close to the previous GGA studies by Hakkinen *et al* [10].

Neutral Au trimer (Au$_3$) has a zigzag ground-state geometry with the bonds of length 2.581 Å at an angle of 137.7°. The Hirshfeld charge analysis shows that the central Au atom has a small positive charge +0.01e. However, the spin density is not so evenly distributed. The unpaired spin population on the central atom is -0.23e compared to -0.38e of the edge atoms, in excellent agreement with -0.39e from an electron spin resonance spectroscopy [11]. The anion trimer (Au$_3^-$) has a linear ground-state geometry with a bond length of 2.600 Å, and the cation trimer (Au$_3^+$) is an equilateral triangle of bond length 2.657 Å. For the anion, the Hirshfeld charge of the central atom is -0.22e, in contrast to the charge -0.39e each of the edge atoms. However, the Hirshfeld charge is evenly distributed for the highly symmetric cation. For both the anion and the cation, there exists no spin fluctuation.

The ground-state neutral Au tetramer (Au$_4$) is a rhombus with bond lengths of 2.706 Å at an acute angle of 58.68°. The geometry is quite close to two equilateral triangles joined together. The Y-shaped conformer is stable but higher in energy (18 meV/atom), not accidentally degenerate with the rhombus as was predicted by a previous plane wave density functional theory calculation with the Becke-Lee-Yang-Parr exchange-correlation functional [12]. The atoms at acute vertices have Hirshfeld charges of -0.08e, and those at obtuse vertices +0.08e. The anion tetramer (Au$_4^-$) is a ziazag chain with the outer bonds of 2.610 Å and the central bond of 2.616 Å, joined at an angle of 157.1°. The inner atoms have Hirshfeld charge of -0.30e each, and the outside atoms -0.20e each. Compared to the zigzag trimer anion, the charge is distributed more evenly and the bond angles are closer to the straight line. The cation tetramer ($Au_4^+$) is a rhombus with bond lengths of 2.676 Å at an acute angle of 61.89°. The bond lengths are smaller than those of the neutral tetramer, and the bond angle slightly larger. The Hirshfeld charge is +0.23e for the atoms at the acute vertices, and +0.27e for the atoms at the obtuse vertices.

Neutral Au pentamer ($Au_5$) has a capped rhombus structure ($C_2$ symmetry) as the ground-state geometry. Au pentamer anion again has a zigzag structure as the ground-state, but without symmetry. Au pentamer cation has a ground-state geometry that looks like a bowtie.

Au hexamer and the ions ($Au_6$, $Au_6^-$, $Au_6^+$) have two-dimensional triangular ground-state structures consisting of four smaller triangles. The neutral and anionic cluster have a higher symmetry ($C_{3v}$), but the cationic cluster is distorted to a $C_2$ symmetry.

Au heptamer ($Au_7$) has a centered hexagonal structure with a distortion that reduces the symmetry to $C_{2v}$. The anion ($Au_7^-$) has a 3-dimensional structure that consists of $Au_6^-$ structure bent and capped by an Au atom in the z-direction. The symmetry is close to $C_{3v}$. The heptamer cation ($Au_7^+$) is a centered equilateral hexagon with $C_{6v}$ symmetry.

The ground-state geometry of Au octamer and its anion ($Au_8$, $Au_8^-$) consists of four triangle joined to form a square at the center. The triangles are close to being equilateral and the overall symmetry is $C_{4v}$. In the case of the cation ($Au_8^+$), the triangles are further distorted from being equilateral and they join to form a rhombus, instead of a square, leading to a $C_{2v}$ symmetry.

The most conspicuous tendency in the structure of small Au clusters is the outstanding stability of triangular subunits and their two-dimensional combinations and arrangements in the neutral and cationic clusters. The lowest energy geometries of neutral and cationic clusters up to eight atom clusters were basically two-dimensional combinations and arrangements of triangular subunits. However, anionic clusters exhibited a different tendency. Up to $Au_5^-$ cluster, zigzag structures were favored. Their geometries resemble those of the neutral and cationic clusters after that. Au clusters up to 8 atoms, regardless of their charge state, favor one-dimensional or two-dimensional structures except the three-dimensional $Au_7^-$ cluster, which is basically $Au_6^-$ structure capped with an Au atom in the z-direction. The ground-state geometries of the neutral Au clusters are summarized in Fig. 1, the anion clusters in Fig. 2, and the cation clusters in Fig. 3.

The binding energies of the Au clusters and the ions are shown in Fig. 4. The binding energies of $Au_n^+$ cluster are calculated by $E_b(Au_n^+) = E(Au_n^+) - (n-1)E(Au) - E(Au^+)$. Even-numbered neutral clusters and odd-numbered ions are more stable with an exception of $Au_7^-$ cluster that has a 3-dimensional structure. Especially, $Au_4$, $Au_6$, $Au_3^-$, $Au_5^-$, and $Au_3^+$ clusters show outstanding stability. The even-odd alternation is also clearly seen in the HOMO-LUMO gap of Au clusters and ions (Fig.5).

We also calculated ionization potentials (Fig. 6) and electron affinities (Fig. 7). The adiabatic ionization energy (or adiabatic electron affinity) of $Au_n$ cluster is calculated by comparing the total-energy of the fully relaxed $Au_n^+$ (or $Au_n^-$) cluster with the total-energy of the neutral $Au_n$ cluster. The vertical ionization energies and the vertical electron affinities are calculated without relaxing the cluster geometry to simulate the Franck-Condon principle. Our results compare well with experiments [9, 12-15] and previous theoretical studies [12, 16]. The vertical ionization energies exhibit the even-odd alternation observed in the experiment [9] and stay close to the experimental values.

To summarize, we have determined the equilibrium structures of small Au clusters and their anions and cations up to 8 atoms and calculated the ionization energies, the electron affinities, and the HOMO-LUMO gaps. We find a large Hirshfeld charge variation in the case of the linear $Au_3^-$ cluster. The ground-state geometries are one- or two-dimensional, the only exception being the three-dimensional $Au_7^-$. The ionization energies and electron affinities of Au clusters are in reasonably good agreement with available experiments.

## ACKNOWLEDGMENTS

The authors are much indebted to the librarians of Daegu University and wish to express special thanks for their great service. The authors also wish to thank the Physics Department of Daegu University for kindly permitting us to form Beowulf Clusters out of the Physics Computer Lab. This research was supported in part by the Daegu University Research Funds.

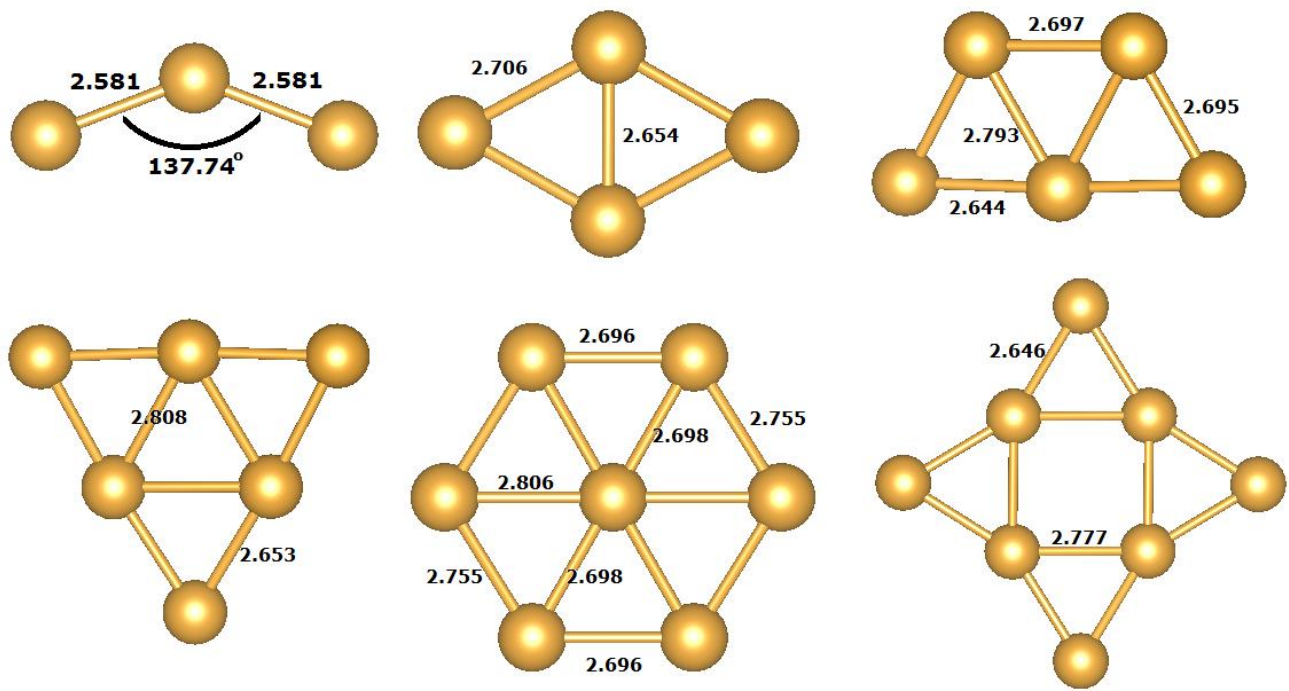

Fig. 1. The most stable geometries of the neutral Au clusters.

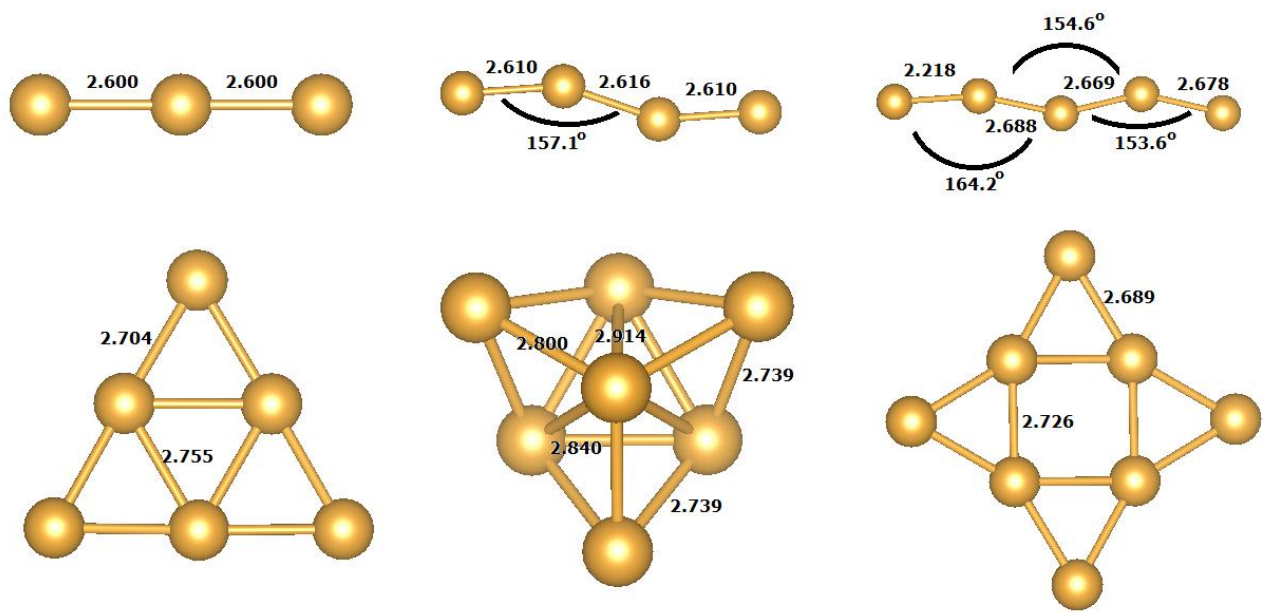

Fig. 2. The most stable geometries of the anionic Au clusters.

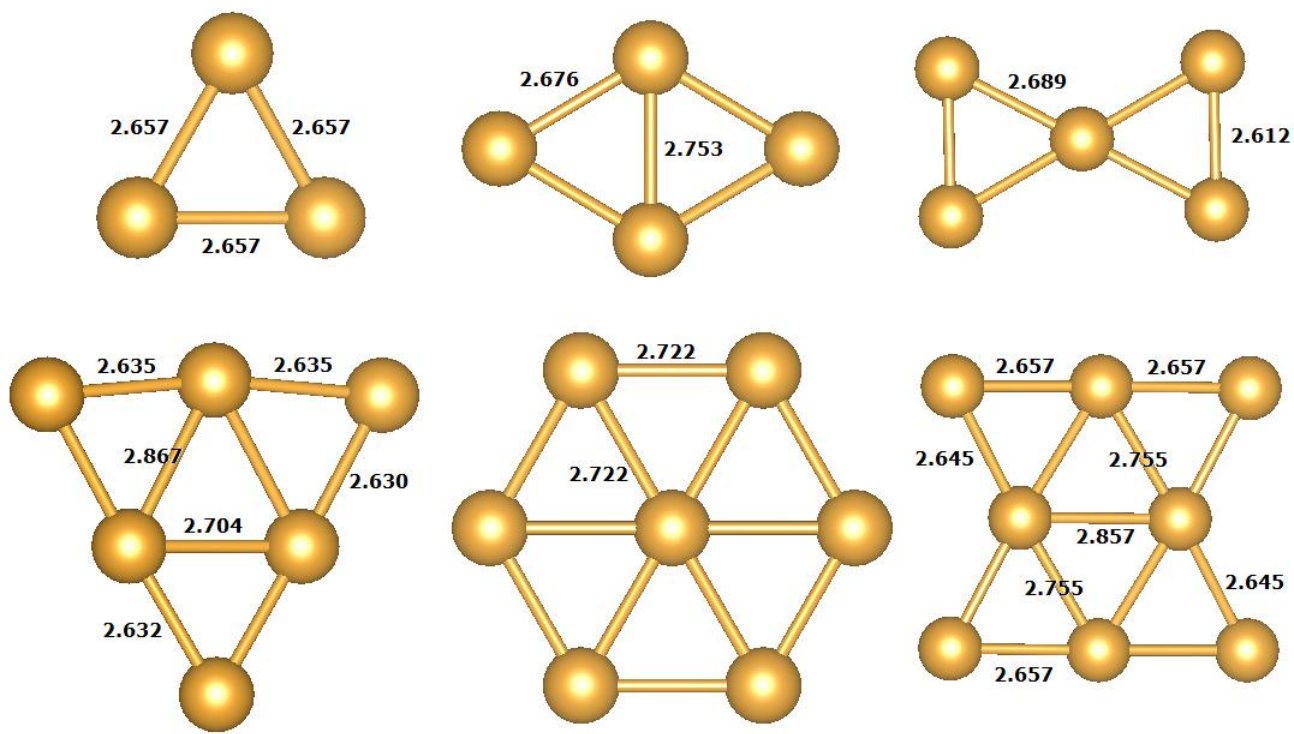

Fig. 3. most stable geometries of the cationic Au clusters.

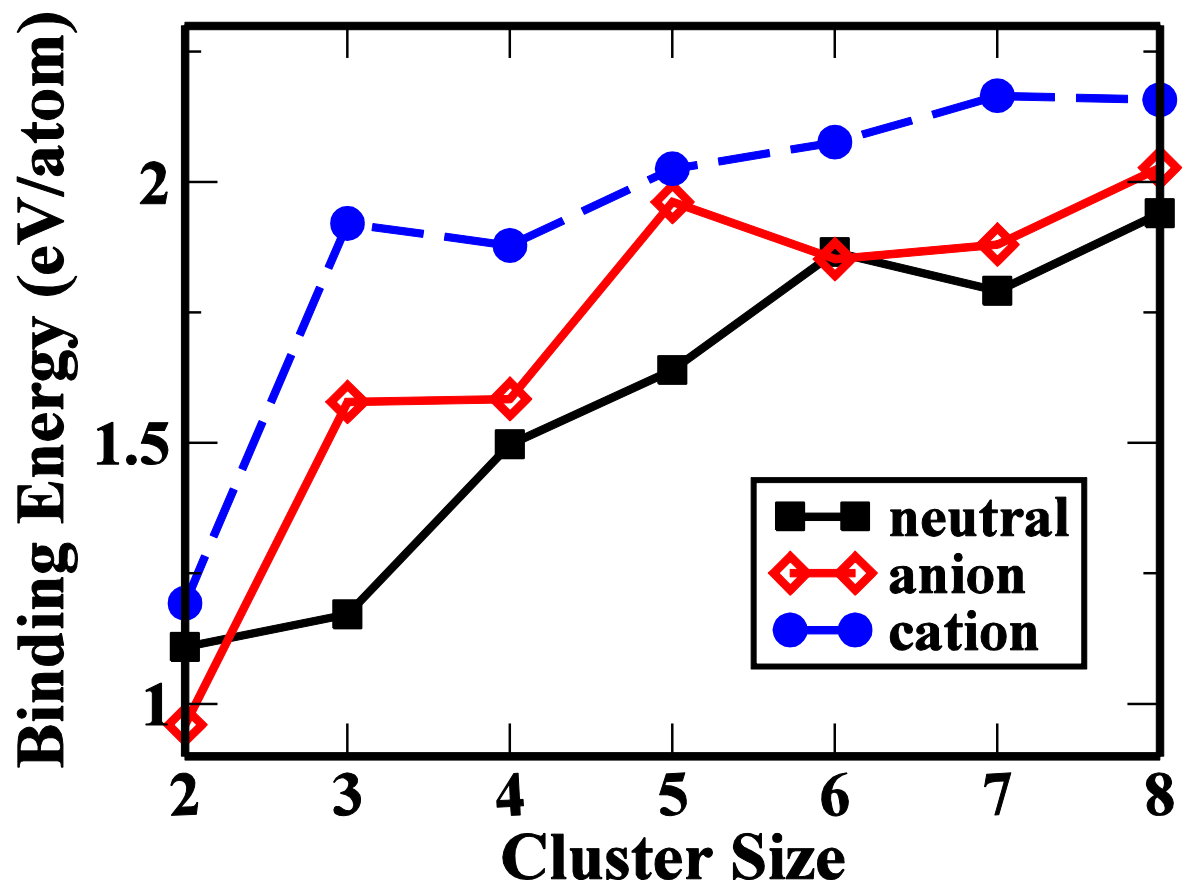

Fig. 4. The binding energy [eV] as a function of the cluster size.

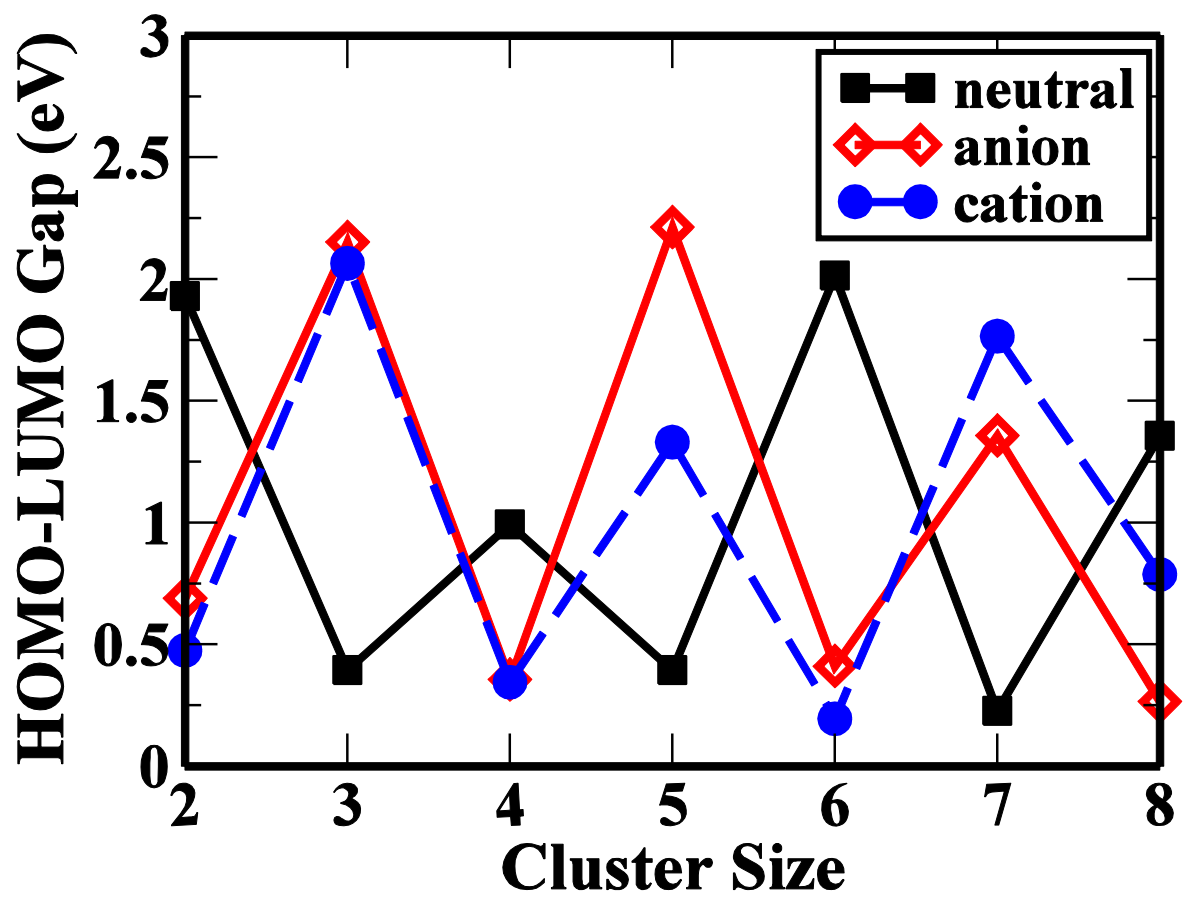

Fig. 5. The HOMO-LUMO gap [eV] as a function of the cluster size.

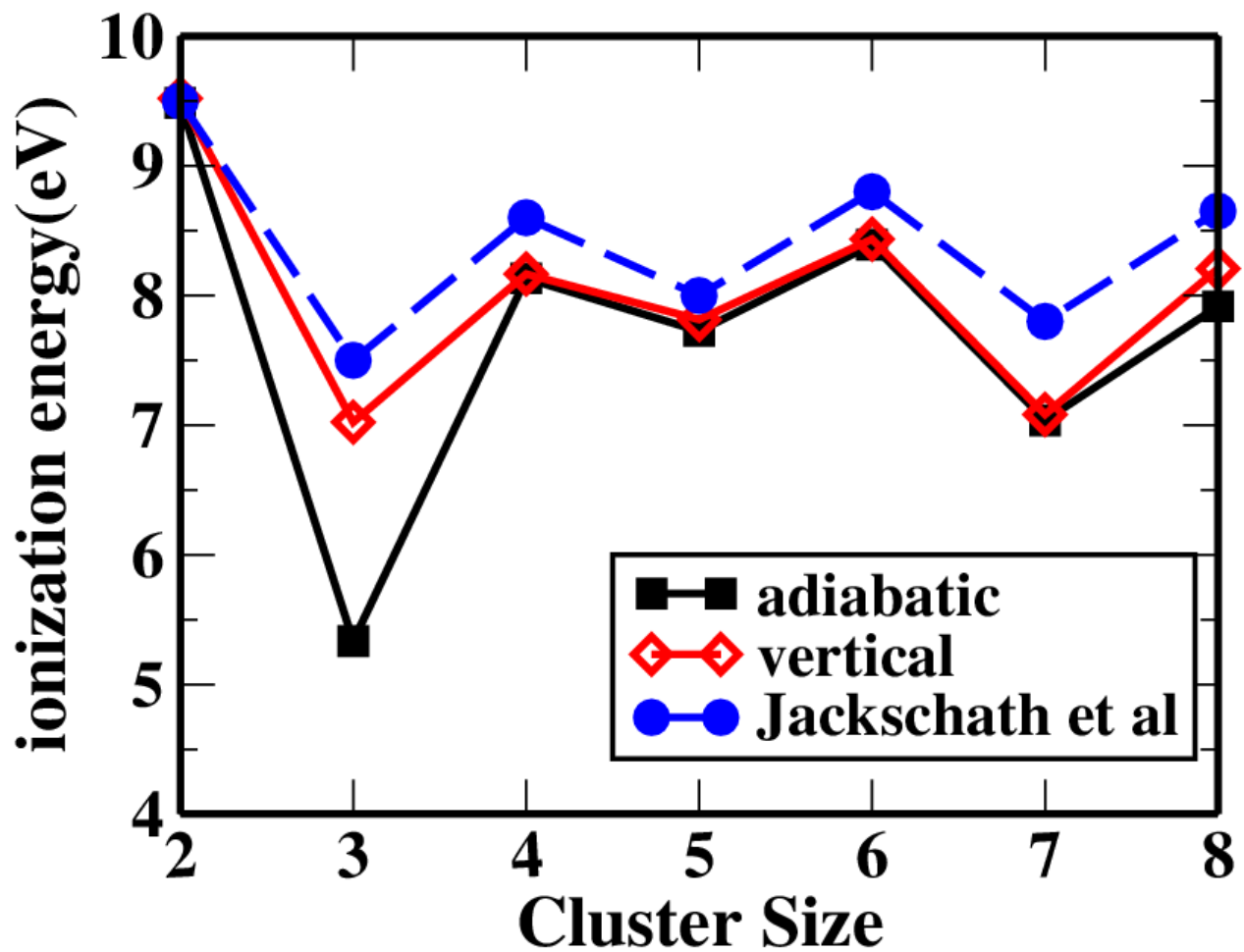

Fig. 6. The ionization energies [eV] as a function of the cluster size.

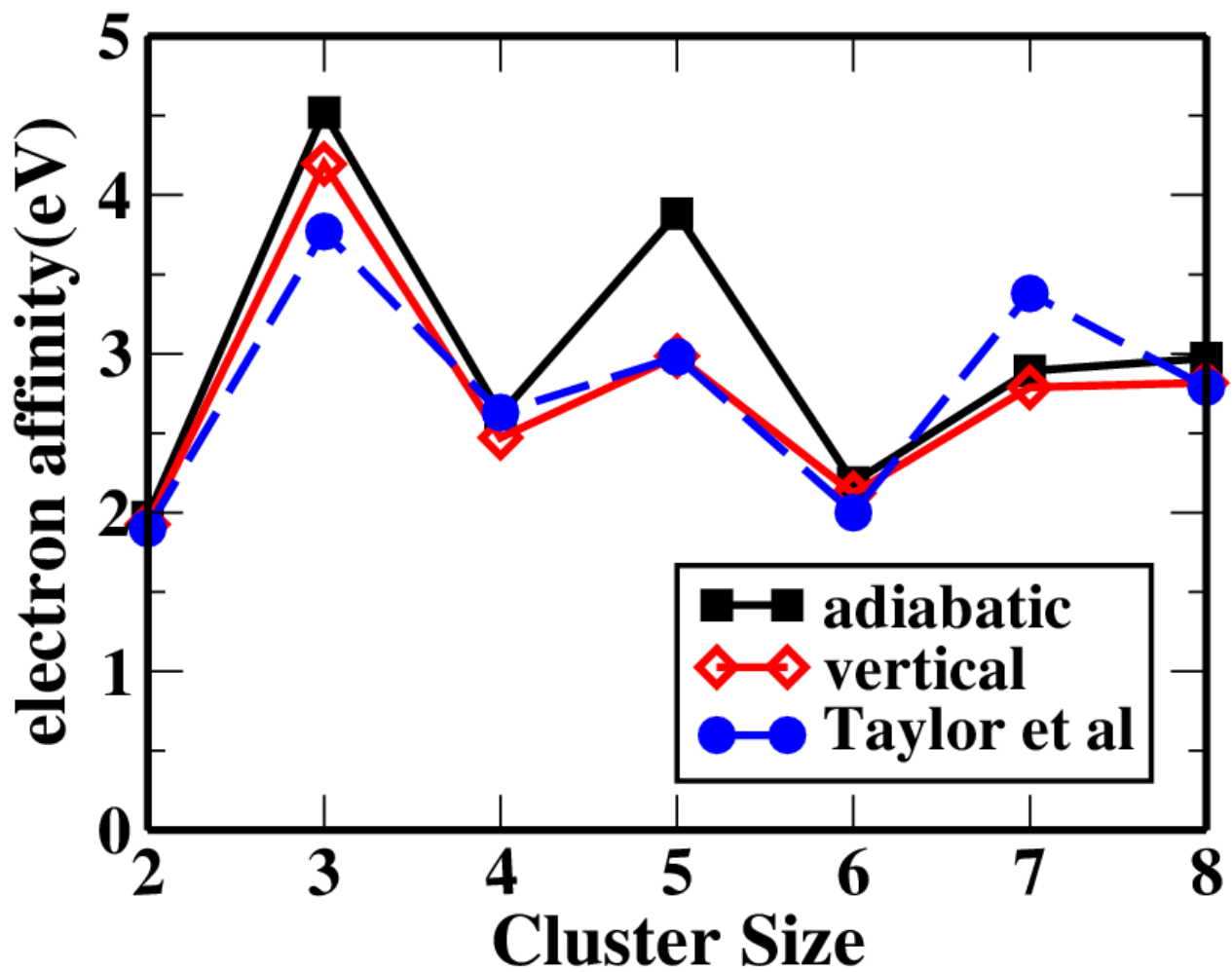

Fig. 7. The electron affinities [eV] as a function of the cluster size.